\documentclass[aps,prd,twocolumn,showpacs,amsmath,superscriptaddress,amssymb,floatfix,nofootinbib,10pt]{revtex4-1}
\usepackage[colorlinks=true, citecolor=blue, linkcolor=blue, urlcolor=blue]{hyperref}
\usepackage{color,xcolor,graphicx,amsfonts,multirow,amsmath,multirow}
\usepackage{ulem}

\allowdisplaybreaks

\begin{document}

\title{Footprint in fitting $B\to D$ vector form factor and determination for $D$-meson leading-twist LCDA}

\author{Sheng-Bo Wu}
\author{Hai-Jiang Tian}
\author{Yin-Long Yang}
\address{Department of Physics, Guizhou Minzu University, Guiyang 550025, P.R.China}
\author{Wei Cheng}
\address{School of Science, Chongqing University of Posts and Telecommunications, Chongqing 400065, China}
\author{Hai-Bing Fu}
\email{fuhb@gzmu.edu.cn}
\author{Tao Zhong}
\address{Department of Physics, Guizhou Minzu University, Guiyang 550025, P.R.China}
\address{Institute of High Energy Physics, Chinese Academy of Sciences, Beijing 100049, P.R.China}

\begin{abstract}
In this paper, we fit the $B\to D$ vector transition form factor (TFF) by using the data measured by \textsl{BABAR} and Belle Collaborations within Monte Carlo (MC) method. Meanwhile, the $B\to D$ TFF  is also calculated by using the QCD light-cone sum rules approach (LCSRs) within right-handed chiral current correlation function. In the TFF, $D$-meson leading-twist light-cone distribution amplitude (LCDA) is the most important non-perturbative input parameter, the precise behavior of which is mainly determined by the parameter $B_{2;D}$ in the light-cone harmonic oscillator model. Through fitting the TFF, we determine the value $B_{2;D}=0.445$. Then, we present the curve of $D$-meson leading-twist LCDA in comparison with other theoretical approaches. Subsequently, the $B\to D$ TFF $f_{+}^{BD}(q^2)$ at the large recoil region is $f_{+}^{BD}(0)=0.648_{-0.063}^{+0.067}$, which is compared in detail with theoretical estimates and experimental measurements. Furthermore, we calculated the decay width and branching  fractions of the Cabibbo-favored semileptonic decays $B\to D\ell^+\nu_{\ell}$, which lead to the results $\mathcal{B}(B^0\to D^-\ell ^+\nu _{\ell}) =(2.10_{-0.38}^{+0.44})\times 10^{-2}$ and $\mathcal{B}(B^+\to \bar{D}^0\ell ^+\nu _{\ell}) =(2.26_{-0.41}^{+0.48})\times 10^{-2}$. Finally, we predict the CKM matrix element with two scenarios $|V_{cb}|_{\rm SR}=41.55_{-4.50}^{+4.91}\times 10^{-3}$ and $|V_{cb} |_{\rm MC}=41.47_{-2.66}^{+2.55 }\times 10^{-3}$ from $B^0\to D^-\ell^+\nu_{\ell}$, $|V_{cb}|_{\rm SR}=40.54_{-3.80}^{+4.33}\times 10^{-3}$ and $|V_{cb} |_{\rm MC}=40.46_{-1.38}^{+1.40}\times 10^{-3}$ from $B^+\to \bar{D}^0\ell^+\nu_{\ell}$ which are in good agreement with theoretical and experimental predictions.
\end{abstract}
\date{\today}

\maketitle

\section{Introduction}

Heavy flavor physics, as an important field of particle physics, contains rich physical information and plays a crucial role in testing, verifying, and refining the Standard Model (SM). There are many important decay processes in heavy flavor physics that are worth in-depth study, among which the $B \to D$ semileptonic decay is a key process. A precise study of this process can not only accurately determine the Cabibbo-Kobayashi-Maskawa (CKM) matrix element $|V_{cb}|$, but also potentially reveal new physics (NP) effects, thereby further consolidating its central role in both theoretical and experimental physics. Currently, the determination of SM parameters still faces significant precision limitations. For example, the relative uncertainty of $|V_{cb}|$~\cite{Cabibbo:1963yz, Kobayashi:1973fv, Beneke:2000wa, Blake:2016olu, Watanabe:2023zyy} in the second row of the CKM matrix is about $3.4\%$, which is much higher than that of other elements in the same row, such as $|V_{cd}|$ and $|V_{cs}|$. On the other hand, transition form factors (TFFs) are key components in calculating the semileptonic $B\to D$ decay process and typically need to be calculated using phenomenological methods.

For this point, the CLEO~\cite{CLEO:1997yyh,CLEO:1998qvx}, \textsl{BABAR}~\cite{BaBar:2009zxk,BaBar:2023kug}, and Belle~\cite{Belle:2001gkd,Belle:2015pkj,Belle-II:2022ffa} Collaborations have carried out precise measurements of the physical observables in the $B\to D$ decay process, such as TFF, decay branching  fractions, and CKM matrix elements. In order to systematically cross-check their findings, we have summarized the key quantities measured by these experiments in Table~\ref{Tab:I}, along with some theoretical predictions, to facilitate comparison and analysis.

\begin{table*}
\footnotesize
\caption{The observables predicted by experiments and theories for $B\to D \ell^+\nu_\ell$ decays include the branching fractions, the TFFs, and the value of $|V_{cb}|$.}. \label {Tab:I}
\begin{tabular}{lllll}
\hline
Experiments~~~~~~~ &Branching  fraction $\times10^{-2}$~~~~~~~~ & $\mathcal{G}(1)|V_{cb}|$ $\times10^{-2}$~~~~~~~~~~~~&$|V_{cb}|$ $\times10^{-3}$\\
\hline
&$\mathcal{B}_{B^0}=1.87\pm0.15\pm0.32$
\\
\raisebox{2.0ex}[0pt]{CLEO'97~\cite{CLEO:1997yyh}}& $\mathcal{B}_{B^+}=1.94\pm0.15\pm0.34$ & \raisebox{2.0ex}[0pt]{$3.37\pm0.44\pm0.48_{-0.12}^{+0.53}$} & \raisebox {2.0ex}[0pt]{$34.4\pm0.045\pm0.049_{-0.012}^{+0.054}\pm0.025$}\\[1.5ex]

&$\mathcal{B}_{B^0}=2.20\pm0.16\pm0.19$
\\
\raisebox {2.0ex}[0pt]{CLEO'99~\cite{CLEO:1998qvx}}&$\mathcal{B}_{B^+}=2.32\pm0.17\pm0.20$ &\raisebox{2.0ex}[0pt]{$4.48\pm0.61\pm0.37$}&\raisebox{2.0ex}[0pt]{/} \\[1.5ex]

&$\mathcal{B}_{B^0}=2.23\pm0.11\pm0.11$
\\
\raisebox{2.0ex}[0pt]{\textsl{BABAR}'10~\cite{BaBar:2009zxk}}&$\mathcal{B}_{B^+}=2.31\pm0.08\pm0.09$& \raisebox{2.0ex}[0pt]{$4.3\pm0.19\pm0.14$} & \raisebox {2.0ex}[0pt]{$38.47\pm0.09$}\\[1.5ex]

&$\mathcal{B}_{B^0} = 2.15\pm0.11\pm0.14$ &/ &$40.02\pm1.76$
\\
\raisebox{2.0ex}[0pt]{\textsl{BABAR}'23~\cite{BaBar:2023kug}}&$\mathcal{B}_{B^+}= 2.16\pm0.08\pm0.13$ & / &$38.67\pm1.41$\\[1.5ex]

Belle'02~\cite{Belle:2001gkd} &$\mathcal{B}_{B^0}=2.13\pm0.12\pm0.39$ & $4.11\pm0.41\pm0.52$ & $41.9\pm0.045\pm0.053\pm0.030$ \\[1.5ex]

Belle'16~\cite{Belle:2015pkj} &$\mathcal{B}_{B^0}=2.31\pm0.03\pm0.11$ & $4.201\pm0.129$ & $40.83\pm1.13$ \\[1.5ex]

&$\mathcal{B}_{B^0}=1.99\pm0.04\pm0.08$ &/&$38.28\pm1.16$
\\
\raisebox{2.0ex}[0pt]{Belle'22~\cite{Belle-II:2022ffa}} &$\mathcal{B}_{B^+}=2.21\pm0.03\pm0.08$ & / & $38.28\pm1.16$ \\

\hline
Theories~~~~~~~~~~~~~~~~~~ &Branching fraction $\times10^{-2}$~~~~~~~~~~~~~~~~~~~ & $f^{BD}_+(0)$~~~~~~~~~~~~~~~~~~~~~~~~~~~&$|V_{cb}|$ $\times10^{-3}$\\
\hline
&$\mathcal{B}_{B^0}=2.03^{+0.92}_{-0.70}$\\
\raisebox {2.0ex}[0pt]{pQCD'14~\cite{Fan:2013qz}}&$\mathcal{B}_{B^+}=2.19^{+0.99}_{-0.76}$ & \raisebox {2.0ex}[0pt]{$0.52^{+0.12}_{-0.10}$} & \raisebox {2.0ex}[0pt]{/} \\[1.5ex]
LCSR'22~\cite{Gao:2021sav} &/ & $0.552\pm0.216$ & $40.2^{+0.6}_{-0.5}|_{\rm th}\pm{1.4}|_{\rm exp}$ \\[1.5ex]
LQCD'15~\cite{MILC:2015uhg} &/ & $0.672\pm0.027$ & $39.6\pm1.7\pm0.2$ \\
\hline
\end{tabular}
\end{table*}

Theoretically, numerous effective approaches exist to make reasonable predictions for TFF of the $B\to D\ell^+\nu_\ell$ semileptonic decay process, such as the perturbative QCD (pQCD)~\cite{Fan:2013qz, Fan:2015kna, Kurimoto:2002sb}, the QCD light-cone sum rule (LCSRs)~\cite{Zuo:2006dk, Zuo:2006re, Fu:2013wqa, Zhang:2017rwz, Wang:2017jow, Zhong:2018exo, Gao:2021sav}, and the lattice QCD (LQCD)~\cite{MILC:2015uhg, deDivitiis:2007otp, Okamoto:2004xg, Hashimoto:1999yp, Bernard:2008dn, Yao:2019vty, Na:2015kha, Kaneko:2019vkx, Martinelli:2021frl,Harrison:2024iad,Martinelli:2021onb}. Different approaches are suitable in different squared momentum transfer $q^2$-region when calculating $B\to D$ TFF. Combining them may offer better results~\cite{Cui:2022zwm, Cui:2023jiw}. The pQCD method can offer a more suitable contribution in the large recoil region, while the LQCD approach is more appropriate for soft regions with large $q^2$. The QCD LCSRs approach involves both the hard and the soft contributions below $\sim 8$ GeV$^2$. It can be clearly seen from the prediction results of some theories summarized and presented in Table~\ref{Tab:I} that there exist large a gap for the vector TFF from different methods, which leads to the first motivation for this paper. Recently, the large momentum effective theory (LaMET) has been integrated with the Lattice techniques to explore the $B$-meson light-cone distribution amplitude (LCDA)~\cite{Hu:2023bba,Wang:2019msf}. It is known that the $B$-meson LCDA assumes a highly significant role in the investigation of $B$-meson decay phenomena, especially for the soft-collinear effective theory (SCET)~\cite{Gao:2019lta,Shen:2020hfq}. When dealing with the $D$-meson LCDA, it will encounter some problems due to the smaller mass of $D$-meson which is close to $\Lambda_{\rm QCD}$. A determined $D$-meson LCDA will be helpful to study the $D$-meson related decay processes which lead to the second motivation of this paper.

Based on the above discussion, the process is worth to be studied due to the result difference between the experimental measurements and the theoretical calculations. The accurate measurement of $|V_{cb}|$ depends heavily on reliable $B\to D$ TFF. The Monte Carlo (MC) method allows us to effectively use a set of data on the $B\to D$ TFF for related calculations. One can fit the TFF data set provided by multiple experimental groups using the MC method and apply the fitted results to determine the relevant decay widths, branching fractions of semileptonic decay $B\to D\ell\nu_\ell$, and the CKM matrix element $|V_{cb}|$, with the aim of comparing both experimental and theoretical results.

After obtaining the resultant fitted data, we shall adopt the improved QCD LCSRs approach ($\it i.e.$ starting from the chiral correlation function) to calculate the $B\to D$ TFF, enabling the comparison between experimental and theoretical results. Meanwhile, the $D$-meson leading-twist LCDA, which serves as a crucial parameter input, is also considered. The $D$-meson leading-twist LCDA is the dominant contribution, while the most uncertain twist-3 contribution can be suitably eliminated by the chiral correlation function. Moreover, as the twist-4 contribution is small in LCSRs, one can furnish a good platform to test the $D$-meson leading-twist LCDA based on the obtained $B\to D$ TFF. Therefore, we present a model for $D$-meson leading-twist LCDA $\phi_{2;D}(x,\mu_0)$, which is constructed by the Brodsky-Huang-Lepage (BHL) prescription~\cite{Wu:2010zc, Wu:2011gf, Huang:2013yya, Huang:1994dy, Wu:2012kw, Cao:1997hw, Wu:2005kq, Wu:2008yr} together with the Wigner-Melosh rotation effect~\cite{Wigner:1939cj,Melosh:1974cu,Kondratyuk:1979gj}. In this work, the model will incorporate three unknown parameters to appropriately describe the behavior of the $D$-meson leading-twist LCDA, which will be discussed in the next section. Among these unknown parameters, there is a free parameter $B_{2;D}$ that governs the longitudinal behavior of the LCDA. To present a better behavior of the $D$-meson leading-twist LCDA, in this manuscript, the data of the $B\to D$ TFF obtained by fitting with the MC technique are used. Meanwhile, the analytical expression of TFF $f_+^{BD}(q^2)$ obtained by the LCSR method is taken as the fitting object. Finally, the $B_{2;D}$, $A_{2;D}$, and $b_{2;D}$ values of the unknown parameters in the $D$-meson leading-twist LCDA can be explicitly determined. Based on this, the improved LCSRs TFF can be more effectively applied in calculations. Further study, using the $B\to D$ TFFs obtained from the MC and LCSRs techniques, will allow one to assess their behaviors and corresponding physical observables, thereby demonstrating the model's applicability to the $D$-meson leading-twist LCDA.

The remaining parts of paper are organized as follows: In Section~\ref{Sec:2}, we present the theoretical framework for studying the $B\to D$ semileptonic processes. Section~\ref{Sec:3} is dedicated to the numerical analysis, where we provide the TFF, decay branching fractions, and predict the numerical results for the CKM matrix elements. Section~\ref{Sec:4} contains a summary.

\section{Theoretical framework}\label{Sec:2}
The $B\to D$ TFFs $f_+^{BD}(q^2)$ and ${\cal G}(w)$ are defined from the basic matrix element which has the following form
\begin{align}
\langle D(p_D)&|\bar{c}\gamma _{\mu}b|B(p_B)\rangle
\nonumber\\
&=2f_+^{BD}(q^2) p_{D\mu} + [f_+^{BD}(q^2)+f_-^{BD}(q^2)] q_{\mu},
\label{eq:matrix1}
\end{align}
in which $q =(p_B- p_D)$ is the momentum transfer, $p_{D\mu}$ represents the four-momentum of the $D$-meson. Where $v_B={p_B}/{m_B}$ and $v_D={p_D}/{m_D}$, and only the vector current $\bar{c}\gamma _{\mu}b$ contributes to the pseudoscalar-to-pseudoscalar amplitude for the non-perturbative matrix elements $\langle D(p_D)|\bar{c}\gamma_{\mu}b|B(p_B)\rangle$~\cite{Caprini:1997mu}. Meanwhile, it also can be expressed in the following form:
\begin{align}\label{eq:ffv}\nonumber
\langle D(p_D)|\bar{c}\gamma_{\mu}b|B(p_B)\rangle  &=  \sqrt{m_B~m_D}[h_+(w)(v_B+v_D)_\mu
\\
&+h_-(w)(v_B-v_D)_\mu],
\end{align}
where a combination of the functions $h_+(w)$ and $h_-(w)$ gives the $B\to D$ TFF ${\cal G}(w), i.e.$
\begin{equation}
{\cal G}(w)=h_+(w)-\frac{m_B-m_D}{m_B+m_D}h_-(w).
\end{equation}
Moreover, the relation between the two $B\to D$ TFFs ${\cal G}(w)$ and $f_{+}^{BD}(q^2)$, is given by
\begin{equation}
f_{+}^{BD}(q^2)=\frac{m_B+m_D}{2\sqrt{m_B~m_D}}{\cal G}(w) \label{eq:fq+}
\end{equation}

\begin{figure}[t]
\begin{center}
\includegraphics[width=0.45\textwidth]{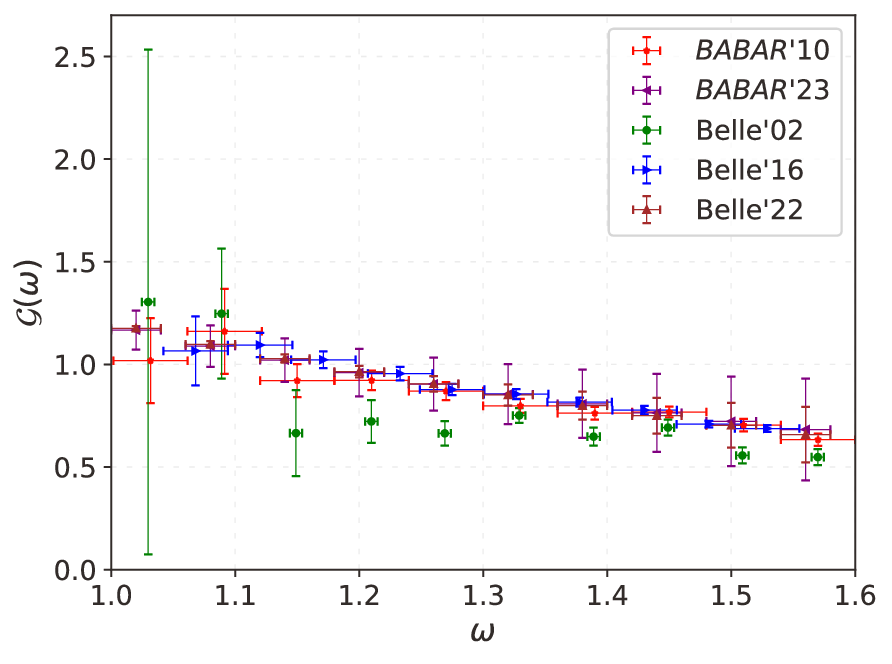}
\end{center}
\caption{The experimental data of TFF $\mathcal{G}(\omega)$ from \textsl{BABAR}'10~\cite{BaBar:2009zxk}, \textsl{BABAR}'23~\cite{BaBar:2023kug}, Belle'02~\cite{Belle:2001gkd}, Belle'16~\cite{Belle:2015pkj} and Belle'22~\cite{Belle-II:2022ffa} collaborations, respectively.}
\label{Fig:G}
\end{figure}

On one hand, \textsl{BABAR} collaboration~\cite{BaBar:2009zxk,BaBar:2023kug} and Belle collaboration~\cite{Belle:2001gkd,Belle:2015pkj,Belle-II:2022ffa} have presented rich datasets for the $B\to D$ vector TFF ${\cal G}(w)$ which are shown in Fig.~\ref{Fig:G}. In order to use this information effectively, one can take the MC approach, which should proceed from a model-independent formula based solely on QCD dispersion relations as proposed by Boyd, Grinstein and Lebed (BGL)~\cite{Boyd:1994tt}.
\begin{equation}
f_{i}(z)=\frac{1}{P_{i}(z)\phi_{i}(z)}\sum_{n=0}^N{a_{i,n}z^n}, ~i=(+,0), \label{eq:fi}.
\end{equation}
The variable $z$ is a function of the recoil variable $w$, which is defined as:
\begin{eqnarray}
    z(w) =\frac{\sqrt{w+1}-\sqrt{2}}{\sqrt{w+1}+\sqrt{2}}
\end{eqnarray}

where $w$ describes the kinematics of the semileptonic decay $B\to D\ell^+\nu_\ell$ and represents the product of the four-velocities of the $B$ and $D$-mesons, namely $w =v_B \cdot v_D$. The $a_{i,n}$ are free parameters, and index $n$ denotes the order at which the series is truncated. Moreover, the $P_{i}(z)$ are the``Blaschke factors" that incorporates the apparent pole in $q^2$ and one can choose $P_{i}(z)$ as stated in Refs.~\cite{Gao:2021sav}. Its definition is as follows
\begin{eqnarray}
P_{+}(z)=\prod_{P_{+}=1}^{3} \frac{z - z_{P_{+}}}{1 - zz_{P_{+}}}
\\
P_{0}(z)=\prod_{P_{0}=1}^{2} \frac{z - z_{P_{0}}}{1 - zz_{P_{0}}}
\end{eqnarray}
Here, the $z_P$ is defined as

\begin{eqnarray}
z_{P}=\frac{\sqrt{t_{+}-m_{P}^{2}}-\sqrt{t_{+}-t_{0}}}{\sqrt{t_{+}-m_{P}^{2}}+\sqrt{t_{+}-t_{0}}}
\end{eqnarray}

Here, $t_+=(m_B+m_D)^2$, and $t_0=t_-=(m_B-m_D)^2$, where the mass $m_P$ refers to the mass of the resonance state of $B_c$. In this study, the information related to the masses of all resonance states, including those with the Blaschke factor, is provided in Refs.~\cite{Dowdall:2012ab,Mathur:2018epb,CMS:2019uhm,Eichten:2019gig}.

The $\phi_{i}(z)$ stands for ``outer functions", which are analytic in the unitary disc $|z|=1$ and are arbitrary, with no poles or branch cuts. Based on this, we adopt the form of the ``outer functions" defined in Ref.~\cite{Bigi:2016mdz}, and their definition is as follows.

\begin{align}\nonumber\label{eq:z1}
    \phi_+ ( z ) =&k_+(1+z)^2( 1-z )^{\frac{1}{2}}
    \\
                  &~~~~~~~~~\times[(1+r)(1-z ) +2\sqrt{r}(1+z )]^{-5},
    \\ \nonumber
    \phi_0 ( z ) =&k_0(1+z)(1-z)^{3/2}
    \\
                  &~~~~~~~~~\times[(1+r)(1-r)+2\sqrt r(1+z)]^{-4},
\end{align}
Where $r =m_D/m_B$, $k_+=12.43$, $k_0=10.11$~\cite{Bigi:2016mdz}. With the choice of the outer function, the coefficients $a_{i,n}$ possess a relatively simple form of unitarity. The coefficient $a_{i,n}$ in Eq.~\eqref{eq:fi} satisfies the condition of unitarity $\sum_{n=0}^N{\left| a_{i,n}\right|^2}\leqslant 1$ for any order $n$. At this stage, the formulaic requirements for performing the MC fitting of $B\to D$ TFF ${\cal G}(w)$ are sufficient.

On the other hand, one can choose the following two-point correlation function from the vacuum to the meson to obtain the improved LCSRs analytical expression of the $B\to D$ vector TFF $f_+^{BD}(q^2)$:
\begin{align}
\Pi_\mu(p,q)&= i\int d^4x e^{ipx} \langle D(p_D)|{\rm {T}}\{\bar{c}(x) \gamma_{\mu} (1+\gamma_5) b(x),
\nonumber\\
&\qquad\times \bar{b}(0)i(1+\gamma_5)d(0)\}|0\rangle \nonumber\\
&= \Pi(q^2,(p_{D}+q)^2) p_{D\mu} + \tilde{\Pi}(q^2,(p_{D}+q)^2)q_\mu .  \label{eq:cc}
\end{align}
Here, we use the right-hand chiral current $\bar{b}(0)i(1+\gamma_5)d(0)$ instead of traditional current $\bar{b}(0)i\gamma_5d(0)$, which can suitably eliminate the uncertainty introduced by the twist-3 LCDAs and make the twist-2 LCDA dominant contribution~\cite{Huang:2001xb}. This can effectively make up for our lack of understanding of the twist-3 wave function and the lack of corresponding $\mathcal{O}({\alpha_s})$ correction. Next, one needs to do a two-step process with Eq.~\eqref{eq:cc}. Firstly, we can insert the complete set of states between the currents in Eq.~\eqref{eq:cc} with the same quantum numbers as $\bar{b}(0)i\gamma_5d(0)$ in the timelike $q^2$-region and isolate the pole term of the lowest pseudoscalar $B$-meson. Then, a threshold $s_0$ will be introduced to separate the ground state $B$ and the excited state $B^{\rm H}$. In the case of the right-hand chiral current, the $B^{\rm H}$ will include the scalar resonances. Therefore, $s_0$ should be set near the lowest scalar $B$-meson. After replacing the contributions form the higher resonances and continuum states with dispersion integrations, the hadronic expression can be obtained. Secondly, in the QCD theory, the correlator can be calculated by carrying out the operator product expansion (OPE) near the small light cone distance $x^2\approx 0$ in the spacelike $q^2$-region, where main nonperturbative inputs are parameterized into the LCDAs of the final meson. Then, with the help of the dispersion relation and quark hadron duality, the OPE can be matched with the hadron expression. In order to remove the subtraction term in the dispersion relation and exponentially suppress the contribution of unknown excitation resonance, one needs to use the Borel transformation. Finally, the LCSRs for the TFF $f^{BD}_+(q^2)$ can be obtained
\begin{align}
f_+^{BD}(q^2)&=\frac{m_{b}^{2}f_D}{m_{B}^{2}f_B}e^{\frac{m_{B}^{2}}{M^2}}\left\{ \int_{\Delta}^1{du\exp \!\! \left[ -\frac{m_{b}^{2}\!-\!\bar{u}(q^2\!-\!um_{D}^{2})}{uM^2} \right]} \right.
\nonumber\\
&
\times \left[ \frac{\phi _{2;D}(u)}{u}  - \frac{8 \, m_{b}^{2}  \hspace{0.02cm}[g_1(u) +\, G_2(u)]}{u^3M^4} \,\,+\, \frac{2 \, g_2(u)}{uM^2} \right]
\nonumber\\
& +\int_0^1 \!\! dv \!\! \int D\alpha _i\frac{\theta (\xi \! - \! \Delta )}{\xi^2M^2}
 \exp \! \!  \left[ -\frac{m_{b}^{2}\!-\!\bar{\xi}(q^2\!-\!\xi m_{D}^{2})}{\xi M^2} \right]
\nonumber\\
&
\times \left[ 2\varphi _{\bot}(\alpha _i)+2\tilde{\varphi}_{\bot}(\alpha _i)-\varphi _{\parallel}(\alpha _i)-\tilde{\varphi}_{\parallel}(\alpha _i) \right] \bigg\},
\label{basicfq2}
\end{align}
in which $\bar{u} = 1 - u$, $\xi = \alpha_1 + v \alpha_3$, $\bar \xi = 1 - \xi$, $G_2(u)=\int_{0}^{u}g_{2}(v)dv $ and the integration lower limit is
\begin{align}
\Delta & = \frac{1}{2m_D^2}\Big[\sqrt{(s_0-q^2-m_D^2)^2+4m_D^2(m_b^2-q^2)}
\nonumber\\
& -(s_0-q^2-m_D^2)\Big],
\end{align}
where $s_0$ is the continuum threshold. Furthermore, the functions $g_1(u), g_2(u), \varphi_{\bot(\|)}(\alpha_i), \tilde{\varphi}_{\bot(\|)}(\alpha_i)$ are the two-particle and three-particle twist-4 LCDAs, respectively, which are similar to the kaonic case with $SU_f(3)$-breaking effect, as introduced in Ref.~\cite{Ball:1998je}. When the masses of the quarks are considered to be infinitely large, our current TFF $f_+^{BD}(q^2)$ aligns with the Isgur-Wise function for TFFs between heavy mesons, as reported in Refs.~\cite{Isgur:1989vq, Isgur:1990yhj}. This indicates that, at least at the leading order, LCSRs for $f_+^{BD}(q^2)$ are equivalent to estimates derived from heavy quark symmetry. However, at the next-to-leading order (NLO) level, the influence of heavy quark masses may introduce discrepancies between these two approaches, which is beyond the scope of this paper. For ease of comparison with experimental analyses found in the literature, we also provide LCSRs for the $B\to D$ TFF within the framework of heavy quark symmetry.

For what concerns the $D$-meson leading-twist LCDA $\phi_{2;D} (x,\mu_0)$, it accounts for the dominant contribution to the $B\to D$ TFF due to the chiral correlation function \eqref{eq:cc}. Therefore, to obtain an accurate behavior of the $D$-meson leading-twist LCDA, a new LCDA is constructed using the light-cone harmonic oscillator (LCHO) model, which has been successfully applied to pseudoscalar mesons $\pi, K$~\cite{Zhong:2021epq, Zhong:2022ecl, Huang:2004su, Tian:2024ubt}, scalar mesons $a_0(980), K_0^*(1430)$~\cite{Wu:2022qqx, Huang:2022xny, Yang:2024ang, Yang:2024jlz}, and vector mesons $\rho, \phi$~\cite{Hu:2024tmc, Fu:2016yzx}, etc. Generally, the $D$-meson leading-twist LCDA $\phi_{2;D} (x,\mu_0)$ and its wave function (WF) $\psi_{2;D} (x,{\bf{k}}_\bot)$ have the definition as follows:
\begin{eqnarray}\label{DA_WF}
\phi_{2;D} (x,\mu_0) = \frac{{2\sqrt 6 }}{{f_D }} \int_{|{\bf{k}}_\bot|^2 \le \mu_0 ^2 } {\frac{{d{\bf{k}}_ \bot  }}{{16\pi ^3 }} \psi_{2;D} (x,{\bf{k}}_\bot )},
\end{eqnarray}
where $f_{D}$ is decay constant and ${\bf k}_\bot$ stands for the transverse momentum. For the WF $\psi_{2;D}(x,\mathbf{k}_\bot)$, one can establish a relation between the equal-time WF in the rest frame and light cone WF in the finite momentum frame based on the Brodsky-Huang-Lepage (BHL) description, which can be written as
\begin{eqnarray}
\psi_{2;D}(x,\mathbf{k}_\bot) = \chi_{2;D}(x,\mathbf{k}_\bot)\psi_{2;D}^R(x,\bf{k}_\bot),
\end{eqnarray}
where the total spin-space WF is
\begin{eqnarray}
    \chi_{2;D}(x,\mathbf{k}_\bot)= \frac{\bar{x}m_c+xm_d}{\sqrt{\textbf{k}^2_\perp+(\bar{x}m_c+x m_d)^2}},
\end{eqnarray}
while the spatial WF can be expressed as,
\begin{align} \nonumber
&\psi_{2:D}^R(x, {\bf k}_\bot) =A_{2;D}[1+B_{2;D}C_2^{3/2}(x-\bar x)]
\\
&\qquad \times{\rm exp}\left[-b_{2;D}^2\left(\frac{{\bf k}_\bot^2 +m_c^2}{x}+ \frac{{\bf k}_\bot^2 +m_c^2}{\bar x}\right) \right].
\end{align}

After integrating over the transverse momentum dependence ${\bf k}_\bot$, one can obtain the $D$-meson leading-twist LCDA $\phi_{2;D} (x,\mu_0)$, namely:
\begin{align}
\phi _{2;D}&(x,\mu_0)  = \frac{A_{2;D}\sqrt{6x\bar{x}}\mathrm{Y}}{8\pi ^{3/2}f_D b_{2;D}}[1+B_{2;D}\times C_{2}^{3/2}(x-\bar{x})]
\nonumber\\
& \times
\exp \biggl[ -b_{2;D}^{2}\frac{xm_{d}^{2}+\bar{x}m_{c}^{2}-\mathrm{Y}^2}{x\bar{x}} \biggr]
\nonumber\\
& \times \left[ \mathrm{Erf}\left( \frac{b_{2;D}\sqrt{\mu_0^{2}+\mathrm{Y}^2}}{\sqrt{x\bar{x}}} \right) -\mathrm{Erf}\left( \frac{b_{2;D}\mathrm{Y}}{\sqrt{x\bar{x}}} \right) \right] ,
\label{phi2d}
\end{align}
where $A_{2;D}$ is the normalization, and $b_{2;D}$ is harmonic parameter that dominates the WF's transverse distribution. The $B_{2;D}$ is free parameter, which will be discussed in the following section. The error function ${\rm{Erf}}(x)$ is defined as $ {\rm{Erf}} (x)=2\int^x_0{\exp({-t^2})dt} /\sqrt{\pi} $, ${\rm{Y}} = x m_d+\bar{x}m_c$ and $\bar{x} = (1 - x) $, the $C_{2}^{3/2}(x-\bar{x})$ is the Gegenbauer polynomial. Additionally, after accurately determining the parameters $B_{2;D}$, $A_{2;D}$, and  $b_{2;D}$, to further ensure the accuracy and reliability of the obtained parameters, verification work can be carried out with the help of the following formula:
\begin{itemize}
\item The WF normalization condition:
\begin{eqnarray}
\int^1_0 dx \int \frac{d^2 \textbf{k}_\bot}{16\pi^3} \psi_{2;D}(x,\textbf{k}_\bot) = \frac{f_D}{2\sqrt{6}} \;. \label{Pd_1}
\end{eqnarray}

\item The probability of finding the leading valence-quark state in $D$-meson ($P_D$), which is $\simeq 0.8$~\cite{Li:1988hr,Guo:1991eb,Sun:2010oyk}, as defined in the following form:
    \begin{eqnarray}
     P_D=\int^1_0{dx \int_{|\mathbf{k}_{\perp}|^2\le \mu_0^2} {\frac{d^2 \mathbf{k}_\perp}{16\pi^3}|\psi_{2;D}(x, \mathbf{k}_{\perp} )|^2}}.
     \label{Pd_2}
    \end{eqnarray}
\end{itemize}
The LCWF involved in the above two conditions is constructed based on the BHL concept. Therefore, conducting a reasonable verification of these three parameters can not only test the viability of the model we constructed, but also help in evaluating its rationality, providing crucial evidence for the reliability of the model.

Finally, to calculate the corresponding decay branching fraction, the following formula can be used to define the decay width of the semileptonic decay $B\to D$
\begin{align}
\frac{d\Gamma}{dq^2}&(B{\to}D\ell\bar{\nu}_\ell)  = \frac{{G_F^2 |V_{cb}|^2 }}{192\pi^3 m_B^3} \left(1 - \frac{m_\ell^2}{q^2} \right)^2
\nonumber\\
&\times  \bigg[ \bigg( {1 + \frac{{m_\ell^2 }}{2q^2}} \bigg)\lambda ^{\frac{3}{2}} (q^2) |f_{+}^{BD}(q^2 )|^2
\nonumber\\
&
+ \frac{{3m_\ell^2 }}{{2q^2 }}(m_B^2  - m_D^2 )^2 \lambda ^{\frac{1}{2}} (q^2 ) |f_{0}^{BD}(q^2)|^2 \bigg] , \label{Vcb1}
\end{align}
Here, the Fermi constant $G_F=1.166\times10^{-5} \;{\rm GeV^{-2}}$, and the phase space factor $\lambda(q^{2}) = (m_B^2  + m_D^2  - q^2)^2 - 4 m_B^2 m_D^2$.
When the lepton mass $m_\ell$ is ignored, ${\it i.e.}, m_\ell\to 0$, the decay width will only have the contribution from the vector TFF $f_{+}^{BD}(q^2)$.

\section{Numerical results and discussions}\label{Sec:3}
In order to carry out the phenomenological analysis, the following input parameters are adopted: for the $b, c$-quark masses, we take $m_b =4.78\pm 0.06$ GeV and $m_c =1.5\pm 0.05$ GeV, while the masses of $B$-meson and $D$-meson are $m_B =5.279$ GeV and $m_D =1.869$ GeV, as listed by the Particle Data Group (PDG)~\cite{ParticleDataGroup:2024cfk}. For the decay constants of $B$-meson and $D$-meson, we take $f_B=0.19\pm0.013$ GeV and $f_D=0.212\pm0.007$ GeV~\cite{ETM:2016nbo,Bazavov:2017lyh,Dowdall:2013tga,Hughes:2017spc,Carrasco:2014poa}.

\begin{figure}[t]
\begin{center}
\includegraphics[width=0.45\textwidth]{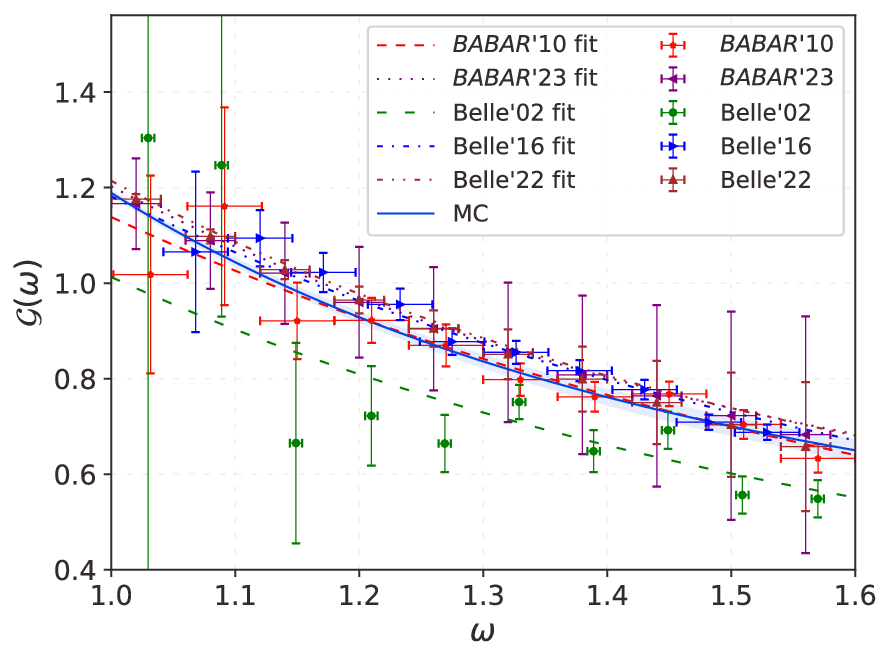}
\end{center}
\caption{The fitting results which are derived from the experimental data points from \textsl{BABAR}'10~\cite{BaBar:2009zxk}, \textsl{BABAR}'23~\cite{BaBar:2023kug}, Belle'02~\cite{Belle:2001gkd}, Belle'16~\cite{Belle:2015pkj}, Belle'22~\cite{Belle-II:2022ffa} collaborations by using the MC method.}
\label{Fig:fq+}
\end{figure}

By extracting the data points of TFF from the \textsl{BABAR} and Belle Collaborations, we employ the MC method for fitting and ensure they are reasonably dispersed throughout the study area, as shown in Fig.~\ref{Fig:fq+}.
The fitting result not only improves the  accuracy but also reflects a more comprehensive connection between $\mathcal{G}(\omega)$ and $\omega$. The description in Fig.~\ref{Fig:fq+} is described in detail below.
\begin{itemize}
\item  In Fig.~\ref{Fig:fq+}, the extracted $B\to D$ TFF data sets from the \textsl{BABAR}~\cite{BaBar:2009zxk,BaBar:2023kug} and Belle~\cite{Belle:2001gkd,Belle:2015pkj,Belle-II:2022ffa} Collaborations are shown, respectively. Meanwhile, we also give the fitted results for each experiment data by using MC method. Moreover, we present the total fitted results of {five} experimental data sets. This result is obtained by taking into account the central values and uncertainties of the five experimental data sets, comprehensively.

\item  By examining the lines in Fig.~\ref{Fig:fq+}, the total fitted result exhibits better behavior compared to other five fitted curves, which may enable more accurate theoretical predictions for the physical observables of the semileptonic decays $B\to D\ell^+\nu_\ell$.
\end{itemize}

\begin{table*}[t]
\begin{center}
\renewcommand{\arraystretch}{1.4}
\footnotesize
\caption{The BGL fitting method is applied to the experimental measurement of TFF data points, and the corresponding truncation parameter $a_{n},n=(0,1,2,3)$ is successfully extracted. In this process, the BGL sequence was truncated at $N = 3$.}
\begin{tabular}{llllll}
\hline
~~~~~~~~~~~~~~~~~~~~~~~&$a_0$~~~~~~~~~~~~~~~~~~~~~~~&$a_1$
~~~~~~~~~~~~~~~~~~~~~~~&$a_2$~~~~~~~~~~~~~~~~~~~~~~~&$a_3$~~~~~~~~~~~~~~~~~~~~~~~~~~&$\chi^2/{\rm ndf}$\\
\hline
\textsl{BABAR}'10~\cite{BaBar:2009zxk}&$0.0121\pm0.0006$&$-0.092\pm0.0018$&$0.3519\pm0.0095$&$-0.3616\pm0.1025$&$5.834/6$\\
\textsl{BABAR}'23~\cite{BaBar:2023kug} &$0.0126\pm0.00006$&$-0.0923\pm0.0022$&$0.3319\pm0.0064 $&$ -0.0739\pm0.0556$&$4.613/6$\\
Belle'02~\cite{Belle:2001gkd}&$0.0109\pm0.0008$&$-0.089\pm0.0016$&$0.340\pm0.0020$&$-0.037\pm0.027$&$9.678/6$\\
Belle'16~\cite{Belle:2015pkj}&$0.0126\pm0.0001$&$-0.094\pm0.003$&$0.34\pm0.04$&$-0.1\pm0.6$&$5.568/6$\\
Belle'22~\cite{Belle-II:2022ffa} &$0.0127\pm0.00001$&$ -0.0936\pm0.0004$&$0.3328\pm0.0069$&$-0.1732\pm0.0886$&$3.275/6$\\
MC &$0.0153\pm0.00008$&$-0.0622\pm0.0031$&$0.3596\pm0.0301$&$-0.2205\pm0.1119$&$73.306/46$\\
\hline
\label{table:II}
\end{tabular}
\end{center}
\end{table*}

\begin{figure}[t]
\begin{center}
\includegraphics[width=0.45\textwidth]{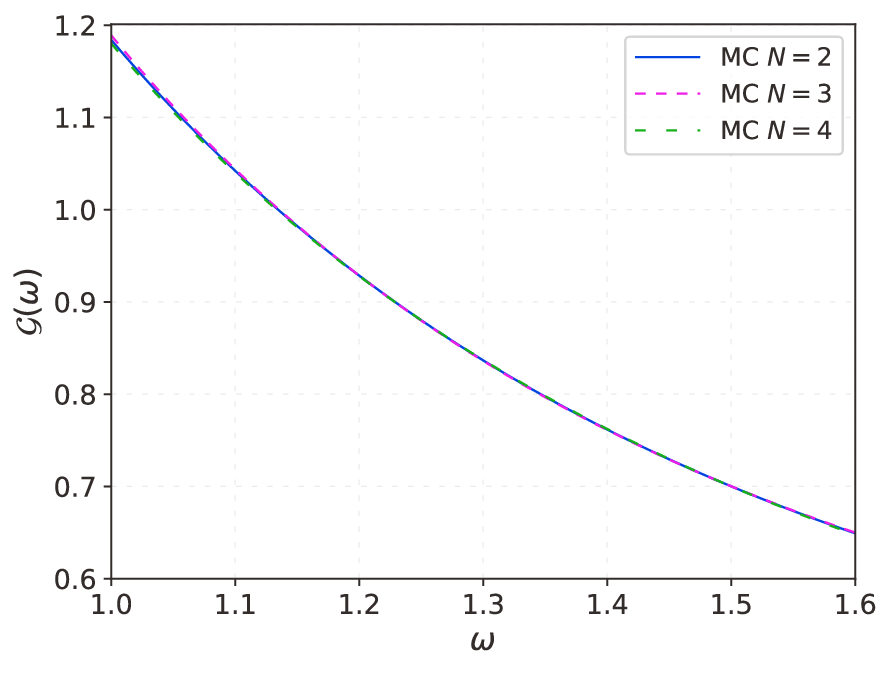}
\end{center}
\caption{{Under the MC method, the fitting results are obtained by comprehensively considering all experimental data points. Additionally, the cases where the BGL series was truncated at the
three orders $N = 2$, $N = 3$ and $N = 4$ were taken into account.} }
\label{Fig:MC}
\end{figure}

To verify the reliability of the results obtained by the MC method, we tested the effects of truncation at $N =(2,3,4)$ during the fitting process. The specific trends are presented in Fig.~\ref{Fig:MC}. It can be seen from this figure that when attempting truncations at $N =2$ and $N=4$, the fitted curves were basically coincident with the curve fitted at $N = 3$. Specifically, the $\chi^2/{\rm ndf}$ values at truncations of $N = 2$ and $N = 4$ are 82.814/47 and 87.591/45 respectively. Therefore, we adopted the truncation strategy of $N=3$ for the subsequent relevant calculations. Moreover, the truncation parameters finally obtained $a_0$, $a_1$, $a_2$, $a_3$, as well as the key indicator $\chi^2/{\rm ndf}$ (chi-squared per degree of freedom, see Ref.~\cite{Franke} used to measure the fitting effect, are all listed in Table~\ref{table:II}. Among $\chi^2/{\rm ndf}$ is an important indicator of fitting quality. When its value is closer to 1, the fitting results are more consistent with the experimental data. It is worth noting that for the fitting results of the five experiments, their $\chi^2/{\rm ndf}$ values are all approach 1. After comprehensively considering all the data points of the TFF, we successfully plotted a curve with a $\chi^2/{\rm ndf}$ value approach 1. The above situation fully demonstrates that the results obtained by the MC method are highly reliable.

Next, for the $B\to D$ TFF calculated by LCSRs, the non-perturbative input parameter twist-2 LCDA $\phi_{2;D}(x,\mu_0)$ needs to be determined, but there remains an undetermined free parameter $B_{2;D}$. At the stage, we shall discuss the $B\to D$ vector TFF $f_+^{BD}(q^2)$ calculated by LCSRs, in which the $D$-meson leading-twist LCDA $\phi _{2;D}(x,\mu_0)$ should be firstly determined, which implicitly finds the three unknown parameters $A_{2;D}, b_{2;D}$, and $B_{2;D}$. In the manuscript, we adopt the BGL MC fit results of  $B\to D$ vector TFF $f_+^{BD}(q^2)$ in the range $q^2\in [0, 7.2]~{\rm GeV^2}$ to determine these parameters. According to the MC fit coefficients $a_0, a_1, a_2$, and $a_3$ in Table~\ref{table:II}, one naturally gets the data of $B\to D$ TFF $f_+^{BD}(q^2)$ and thus we can fit the corresponding $D$-meson leading-twist LCDA $\phi_{2;D}(x,\mu_0)$ via the LCSR analytical expression Eq.~\eqref{basicfq2}. Therefore, the reasonable values of the three parameters are explicitly derived from the BGL MC fit. Before proceeding with the strategy, one needs to determine the continuum threshold $s_0$ and the Borel parameter $M^2$, which can be determined by the self-consistency criteria of the QCD sum rules, $\it i.e.,$

\begin{itemize}
\item To determine the value of $M^2$, we require that the contribution of the continuous states is less than $30\%$ of the total LCSRs;
\item It is also required that the contribution of twist-4 LCDAs does not exceed $5\%$, Therefore, the contribution of twist-4 is neglected;
\item Within the Borel window, the changes of TFFs does not exceed $10\%$;
\item The continuum threshold $s_0$, serving as the boundary between the ground state contributions and higher-mass contributions of $B$-meson, is generally set close to the square of the mass of the first excited state of $B$-meson.
\end{itemize}

\begin{figure}[t]
\begin{center}
\includegraphics[width=0.45\textwidth]{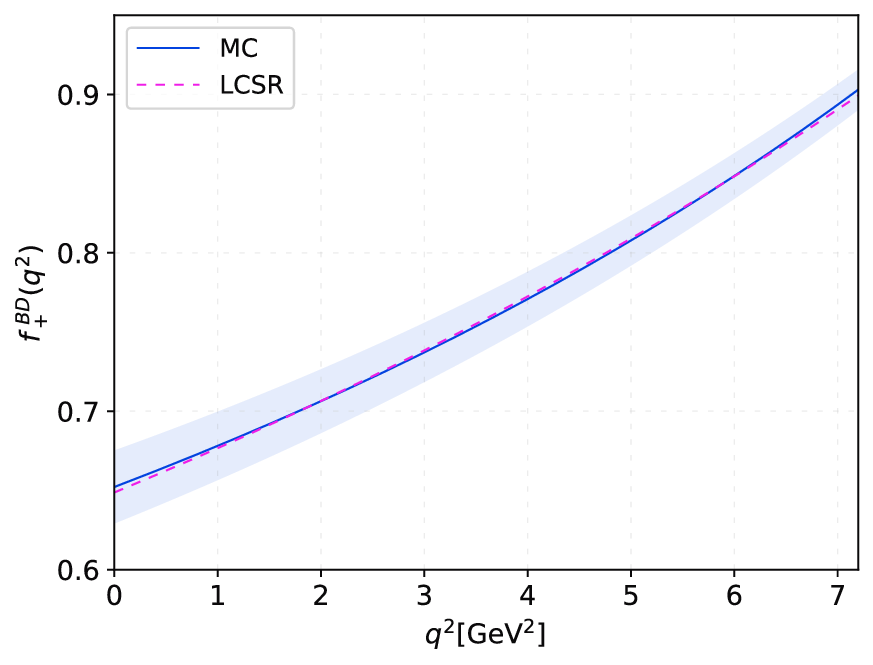}
\end{center}
\caption{The TFFs $f_+^{BD}(q^2)$ versus $q^2$. The solid line is the MC fit and the shadow bands represent errors. The dotted line represents the result of the LCSR.}
\label{Fig:fp}
\end{figure}

After having followed the above criteria, we determined the continuum threshold $s_0=39(1)~\rm{GeV}^2$ and Borel window is taken as $M^2=17(1)~\rm{GeV}^2$. To visually demonstrate how TFF $f_+^{BD}(q^2)$ can be fitted using the above-mentioned method after determining these important input parameters, the following operations were carried out. Given that the LCSRs method is only valid in the low and intermediate $q^2$-regions, in Fig.~\ref{Fig:fp}, we only present the behavior of the TFF of $f_+^{BD}(q^2)$ within the region of $0 \leq q^2\leq 7.2~{\rm GeV^2}$ for both the MC fitting and LCSRs. In Fig.~\ref{Fig:fp}, the solid line represents the MC fitting results, the shaded band indicates the error range, and the dashed line represents the results calculated by LCSR. Through this fitting process, we successfully determined the parameters $B_{2;D}$, $A_{2;D}$, and $b_{2;D}$, and their specific values are listed in Table~\ref{B_results}. Among them, the obtained parameters correspond to the lower limit, the central value, and the upper limit of the $f_+^{BD}(q^2)$ by the MC method, respectively. Based on these parameters, we were able to further explore the behavioral characteristics of the twist-2 LCDA of the $D$-meson.

\begin{table}[b]
\caption{The parameters of the leading-twist LCDA $\phi_{2;D}(x,\mu_0)$ of the $D$-meson at $\mu_0 = 2~\mathrm{GeV}$, which correspond to the lower limit value, the central value and the upper limit value of the $f_+^{BD}(q^2)$ by the MC method, respectively.}
\begin{tabular}{llll}
\hline
&$B_{2;D}$~~~~~~~~~~~&$A_{2;D}$~~~~~~~~~~~&$b_{2;D}$ \\
\hline
Lower~~~~~~~~~~~&$0.2397$ & $9.7736$& $-0.2244$ \\
Center~~~~~~~~~~~&$0.4445$ & $11.7745$ & $-0.2911$ \\
Upper~~~~~~~~~~~&$0.6177$ & $9.0712$& $-0.2657$\\
\hline
\label{B_results}
\end{tabular}
\end{table}

\begin{figure}[t]
\begin{center}
\includegraphics[width=0.45\textwidth]{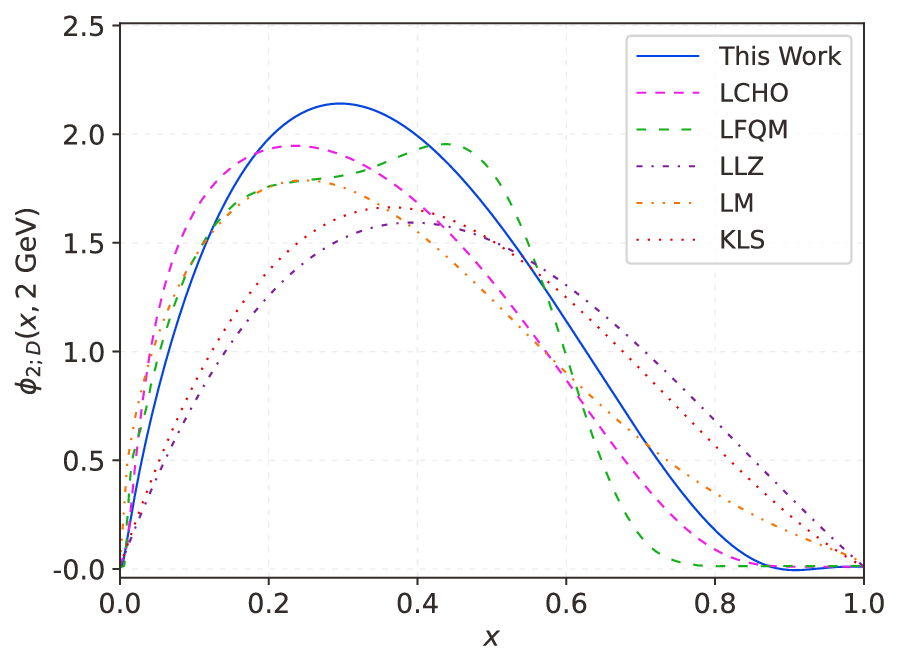}
\end{center}
\caption{Curves of the $D$-meson leading-twist LCDA $\phi _{2;D}(x,\mu_0)$ at $\mu_0=2~\rm GeV$.  The LCDA models in literatures such as KLS model~\cite{Kurimoto:2002sb}, LLZ model~\cite{Li:2008ts}, LM model~\cite{Li:1999kna}, the form with LFQM~\cite{Dhiman:2019ddr}, and  LCHO model~\cite{Zhong:2022ugk} are also shown for comparison.}
\label{Fig:DA1}
\end{figure}

To visually and intuitively demonstrate the specific behavior of the $D$-meson leading-twist LCDA $\phi_{2;D}(x,2~\rm GeV)$, we plot it under different models in Fig.~\ref{Fig:DA1}. And other predictions from the KLS model~\cite{Kurimoto:2002sb}, LLZ model~\cite{Li:2008ts}, LM model~\cite{Li:1999kna}, LFQM model~\cite{Dhiman:2019ddr}, and LCHO model~\cite{Zhong:2022ugk} are also shown for comparison. The detailed descriptions are listed as follows.
\begin{itemize}
\item Apart from our prediction, we also present the LCHO model~\cite{Zhong:2022ugk}, LFQM model~\cite{Dhiman:2019ddr}, LLZ model~\cite{Li:2008ts}, LM model~\cite{Li:1999kna} and KLS model~\cite{Kurimoto:2002sb} results as a comparison in Fig.~\ref{Fig:DA1};
\item As illustrated in Fig.~\ref{Fig:DA1}, it is evident that the tendencies of all the models display a certain degree of consistency. Nevertheless, the $\phi_{2;D}(x,2~\rm GeV)$ model we proposed demonstrates significant advantages in the medium-low $x$ region. Specifically, this model reaches a peak within the range of $x$ approximately from 0.2 to 0.4, which is consistent with the performance of the LM and LCHO models. Nevertheless, it shows a clear difference from the KLS, LLZ, and LFQM models, as the latter three reach their peaks in a relatively higher $x$ range (approximately from 0.3 to 0.5).
\end{itemize}

In order to access  the information of $B\to D$ TFF in the whole kinematic region, we need to extrapolate the LCSRs predictions obtained above toward large momentum transfer with a certain parametrization for TFFs. The physically allowable ranges for the TFFs are $0 \leq q^2\leq q^2_{\rm max} = (m_B - m_D)^2\sim11.64~{\rm GeV^2}$. One can extrapolate it to whole $q^2$-regions via a rapidly $z(q^2,t)$ converging the simplified series expansion (SSE), {\it i.e.}, the TFFs are expanded as~\cite{Bharucha:2015bzk,Bharucha:2010im}:
\begin{eqnarray}
f_+^{BD}(q^2) =\frac{1}{1-q^2/m_{B_{c}^{*}}^2}\sum_{k=0,1,2}{\beta _kz^k( q^2,t_0 )}\label{FSSE}
\end{eqnarray}
where $\beta_k$ are real coefficients and $z(q^2,t)$ is the function,
\begin{eqnarray}
z^k( q^2,t_0 ) =\frac{\sqrt{t_+-q^2}-\sqrt{t_+-t_0}}{\sqrt{t_+-q^2}+\sqrt{t_+-t_0}},
\end{eqnarray}
with $t_{\pm} = (m_{B} \pm m_{D})^2$ and the auxiliary parameter $t_0=t_{\pm}(1-\sqrt{1-t_-/t_+})$. In this approach, the simple pole form $1-q^2/m_{B_{c}^{*}}^2$ can be adopted to characterize low-lying resonance phenomena, where the $m_{B_{c}^{*}}$  stands for $B$-meson resonances. The masses of the low-lying $B$ resonances are mainly determined by their $J^{P}$-states, whose values can be found in Refs.~\cite{Dowdall:2012ab,Mathur:2018epb,CMS:2019uhm,Eichten:2019gig}. The SSE keeps the analytic structure correct in the complex plane and ensures the appropriate scaling, $f_+^{BD}(q^2)\sim 1/q^2$ at large $q^2$. And the extrapolation results can be evaluated by the goodness-of-fit with $\Delta < 1\%$, which is defined as

\begin{eqnarray}
\Delta =\frac{\sum_t{| F_i(t) -F_{i}^{\mathrm{fit}}( t ) |}}{\sum_t{| F_i(t) |}}\times 100.
\end{eqnarray}

\begin{figure}[t]
\begin{center}
\includegraphics[width=0.45\textwidth]{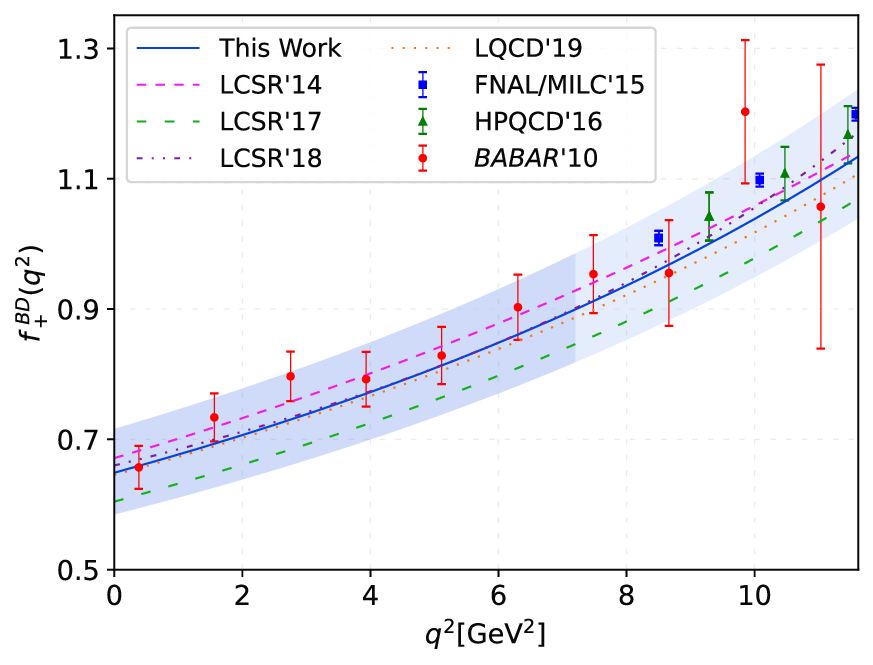}
\end{center}
\caption{{The TFFs $f_+^{BD}(q^2)$ versus $q^2$. The \textsl{BABAR}'10~\cite{BaBar:2009zxk}, LCSR'14~\cite{{Fu:2013wqa}}, LCSR'17~\cite{Zhang:2017rwz}, LCSR'18~\cite{Zhong:2018exo}, LQCD'19~\cite{Yao:2019vty}, FNAL/MILC'15~\cite{MILC:2015uhg} and HPQCD'16~\cite{Na:2015kha} results are present as a comparison.}}
\label{Fig:fp1}
\end{figure}
\begin{table}[b]
\caption {The $B\to D$ TFFs at large recoil point. As a comparison, we also present other predictions.}\label {Tab:II}
\begin{tabular}{ll}
\hline
~~~~~~~~~~~~~~~~~~~~~~~~~~~~~~~~~~~~~~~~~~~~~~~&$f_+^{BD}(0)$\\
\hline
LCSR (This Work) & $0.648_{-0.063}^{+0.067}$\\
MC (This Work) & $0.652\pm0.023$ \\
pQCD'13~\cite{Fan:2013qz}~~&$0.520^{+0.12}_{-0.10}$\\
LCSR'14~\cite{{Fu:2013wqa}}~~&$0.653^{+0.004}_{-0.011}$ \\
LCSR'17~\cite{Zhang:2017rwz}~~~& $0.673^{+0.038}_{-0.041}$\\
LCSR'22~\cite{Gao:2021sav}~~~& $0.552\pm0.216$\\
LCSR'18~\cite{Zhong:2018exo}~~&$0.659^{+0.029}_{-0.032}$\\
LQCD'19~\cite{Yao:2019vty}~~&$0.658\pm0.017$\\
FNAL/MILC'15~\cite{MILC:2015uhg}~~&$0.672\pm0.027$\\
HPQCD'16~\cite{Na:2015kha}~~&$0.664\pm0.034$\\
\hline
\end{tabular}
\end{table}

After making an extrapolation for the TFFs $f_+^{BD}(q^2)$ to the whole physical $q^2$-region. Then, the behaviors of $B\to D$ TFFs in the whole physical region with respect to squared momentum transfer are given in Fig.~\ref{Fig:fp1}.
Then, we also calculate the numerical result of the $B\to D \ell \bar{\nu}_{\ell}$ TFF $f_+^{BD}(q^2)$ at the large recoil region and give its numerical result in Table~\ref{Tab:II} as well as the comparison with other theoretical and experimental groups, which will be described in detail below.
\begin{itemize}
\item In Fig.~\ref{Fig:fp1}, the darker band represents the LCSRs result of our prediction, while the lighter band represents the SSE prediction;

\item In Fig.~\ref{Fig:fp1}, we show the theoretical predictions and experimental results for comparison, such as LCSR'14(17,18)~\cite{Fu:2013wqa,Zhang:2017rwz,Zhong:2018exo}, LQCD'19~\cite{Yao:2019vty}, FNAL/MILC'15~\cite{MILC:2015uhg}, and HPQCD'16~\cite{Na:2015kha}. Upon comparison, it is evident that our results and their associated uncertainties exhibit strong agreement with those from other predictions, thereby reinforcing the applicability of our $D$-meson twist-2 LCDA model.

\item In Table~\ref{Tab:II}, we present the predictions for the $B\to D$ TFF $f_+^{BD}(0)$ at large recoil region, alongside other results for comparison. Upon comparison, we find that our results are consistent with those of other theoretical and experimental groups, falling within the margin of error.
\end{itemize}

\begin{figure}[t]
\begin{center}
\includegraphics[width=0.45\textwidth]{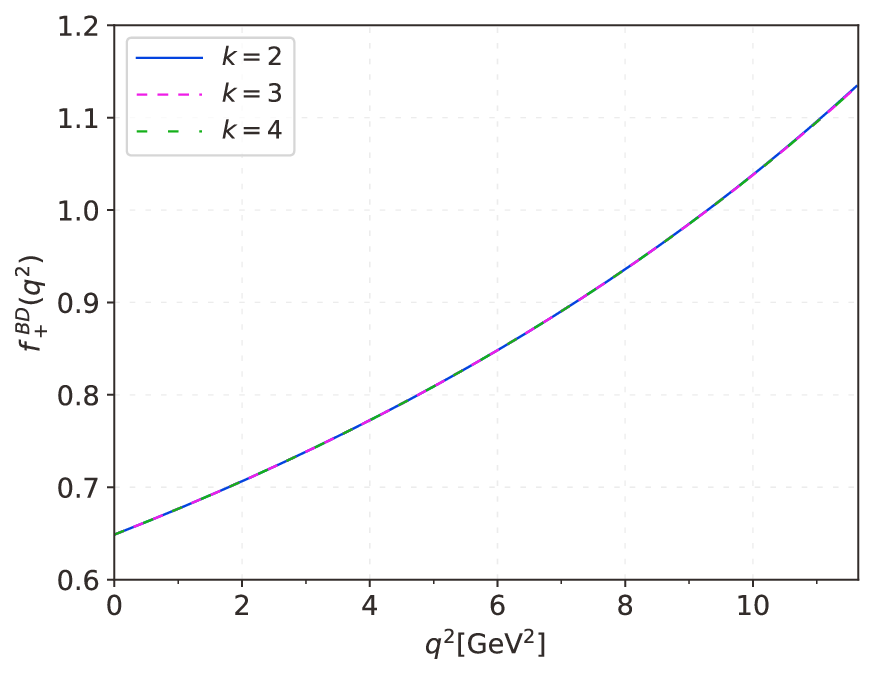}
\end{center}
\caption{The SSE for the TFFs $f_+^{BD}(q^2)$ up to $k=(2,3,4)$ orders, separately.}
\label{Fig:fpk}
\end{figure}

\begin{figure}[t]
\begin{center}
\includegraphics[width=0.45\textwidth]{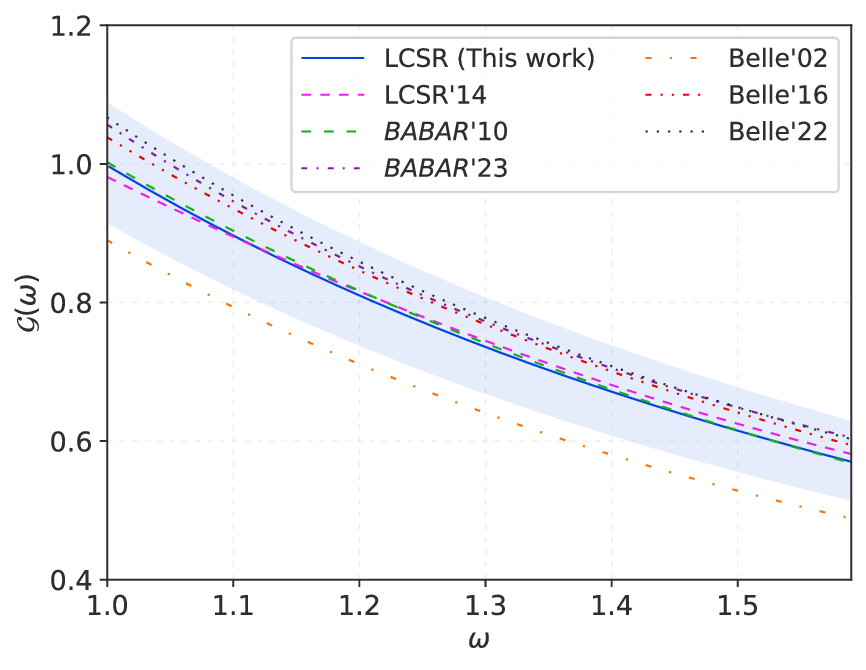}
\end{center}
\caption{{The TFFs ${\cal G}(\omega)$ versus $\omega$. The solid blue line represents the LCSR prediction, and the shaded area represents its error. The predictions of \textsl{BABAR}'10~\cite{BaBar:2009zxk}, \textsl{BABAR}'23~\cite{BaBar:2023kug}, Belle'02~\cite{Belle:2001gkd}, Belle'16~\cite{Belle:2015pkj}, Belle'22~\cite{Belle-II:2022ffa} and LCSR'14~\cite{Fu:2013wqa} are also given as a comparison.}}
\label{Fig:Gw}
\end{figure}

\begin{table}[b]
\caption{The fitted parameters $m_{B_{c}^{*}}$, $\beta_k$ and quality of extrapolation $\Delta$ for the center value $f_+^{BD(\rm C)}(q^2)$, the upper limit $f_+^{BD(\rm U)}(q^2)$ and the lower limit $f_+^{BD(\rm L)}(q^2)$.}
\label{tab:FSSE}
\begin{tabular}{llll}
\hline
~~~~~~~~&$f_+^{BD(\rm C)}(q^2)$~~~~~~~&$f_+^{BD(\rm U)}(q^2)$ ~~~~~~~~&$f_+^{BD(\rm L)}(q^2)$ \\
\hline
$m_{B_{c}^{*}}$~~~~~~~~&$6.332$&$6.332$&$6.332$\\
$\beta_{0}$~~~~~~~~& $0.648$ & $0.715$ & $0.584$ \\
$\beta_{1}$~~~~~~~~& $-2.228$ & $-2.302$ & $-2.149$ \\
$\beta_{2}$~~~~~~~~& $3.107$ & $3.480$ & $3.531$ \\
$\Delta$~~~~~~~~&$0.00051\%$ & $0.00046\%$ & $0.00293\%$ \\
\hline
\end{tabular}
\end{table}

\begin{table}[b]
\caption{Based on the results of $\beta_k$ in Table~\ref{tab:FSSE}, we have provided in detail the specific results of the TFF $f_+^{BD}(q^2)$ at different values of $q^2$.}
\label{tab:Fq2}
\begin{tabular}{llll}
\hline
$q^2(\rm GeV^2)$~~~~~~~~&$f_+^{BD}(q^2)$~~~~~~~~~~~ &$q^2(\rm GeV^2)$ ~~~~~~~~&$f_+^{BD}(q^2)$ \\
\hline
$0$ & $0.648_{-0.063}^{+0.067}$ & $8$ & $0.936_{-0.083}^{+0.088}$ \\
$2$ & $0.706_{-0.068}^{+0.071}$ & $10$ & $1.038_{-0.089}^{+0.096}$ \\
$4$ & $0.772_{-0.072}^{+0.076}$ & $11$ & $1.096_{-0.092}^{+0.101}$ \\
$6$& $0.848_{-0.077}^{+0.082}$ & $11.64$ & $1.135_{-0.094}^{+0.103}$ \\
\hline
\end{tabular}
\end{table}

To demonstrate the rationality of using SSE to extrapolate the TFF $f_+^{BD}(q^{2})$ to the entire $q^{2}$ region, we present the behavior of $f_+^{BD}(q^{2})$ in Fig.~\ref{Fig:fpk} when $k = (2, 3, 4)$ in Eq.~\eqref{FSSE}. It is found that three SSE curves for $f_+^{BD}(q^{2})$ are almost coincident with each other. Additionally, the real coefficients $\beta_{0,1,2}$ and quality of extrapolation $\Delta$ corresponding to the center value $f_+^{BD(\rm C)}(q^2)$, upper limit $f_+^{BD(\rm U)}(q^2)$ and lower limit $f_+^{BD(\rm L)}(q^2)$ are listed in Table~\ref{tab:FSSE}. The uncertainty is mainly due to the input parameters in the LCSR.  Where the values of $\Delta$ for the $B \to D$ TFFs do not exceed $0.00293\%$. Meanwhile, we present the specific results of the TFF $f_+^{BD}(q^{2})$ under different $q^2$-values in Table~\ref{tab:Fq2}. It can be seen that the variation trend after the extrapolation using the SSE method is basically consistent with that of the LCSR in the low and intermediate $q^2$-regions. These analyses fully demonstrates the extremely high feasibility of this method.

In the research framework of $B \to D$ semileptonic decays, the study of TFF $\mathcal{G}(\omega)$ occupies a crucial position. Theoretically, the Heavy Quark Effective Theory (HQET) provides a powerful tool for the calculation of the form factor $\mathcal{G}(\omega)$. By accurately determining $\mathcal{G}(\omega)$ and combining it with the corresponding decay width formula, it is possible to independently extract the CKM matrix element $|V_{cb}|$~\cite{Caprini:1997mu}. As a crucial parameter characterizing the mixing and decay of various quark flavors, the precise measurement of the CKM matrix element is of critical importance for deeply analyzing the SM and its derivative extended theories. Experimentally, $\mathcal{G}(\omega)$ is generally parameterized in the following expansion form:
\begin{align}
{\cal G}_D(w) &= {\cal G}_D(1)\Big[ 1 - {\hat{\rho}} _D^2(w - 1) + {\hat c_D}{(w - 1)^2}
\nonumber\\
& + {\cal O}((w - 1)^3) \Big]. \label{bellefit}
\end{align}

For comparison with our theoretical estimate, we also include the results from the parameterization Eq.~\eqref{bellefit}.

In Fig.~\ref{Fig:Gw}, the MC prediction with its uncertainties are given. The predictions of \textsl{BABAR}'10~\cite{BaBar:2009zxk}, \textsl{BABAR}'23~\cite{BaBar:2023kug}, Belle'02~\cite{Belle:2001gkd}, Belle'16 ~\cite{Belle:2015pkj}, Belle'22~\cite{Belle-II:2022ffa} and LCSR'14~\cite{Fu:2013wqa} are also given as a comparison. Through comparison, it can be found that the variation trend of our results is highly consistent with those predicted by theories and experiments. Moreover, within the allowable error range, our results show a good agreement with both theoretical and experimental predictions.

\begin{figure}[t]
\begin{center}
\includegraphics[width=0.45\textwidth]{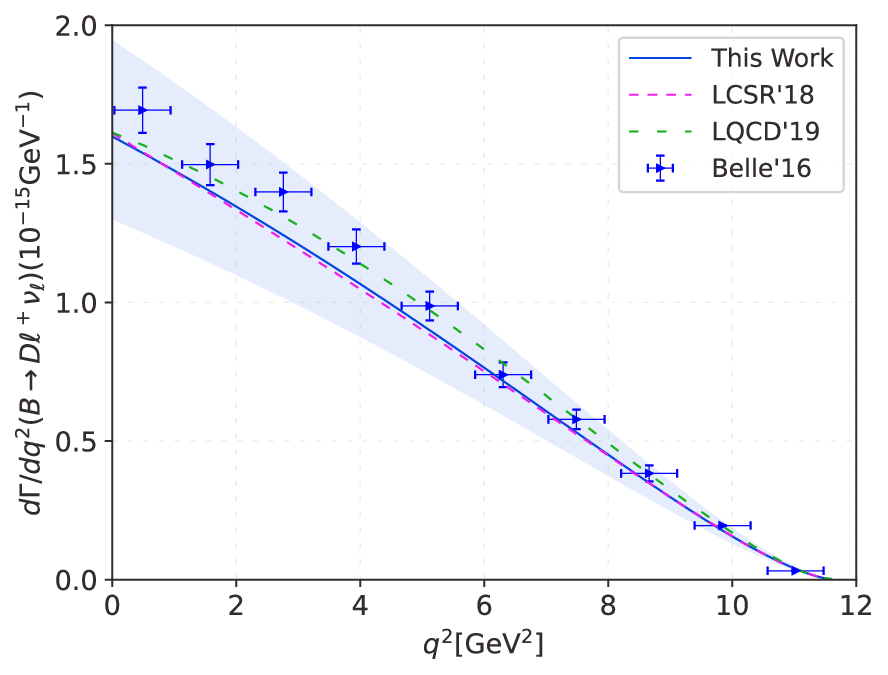}
\end{center}
\caption{{The differential decay width for $B\to D\ell^+\nu_{\ell}$ within uncertainties (in unit: $10^{-15}$). As a comparison, we also show Belle'16 data~\cite{Belle:2015pkj}, as well as the
predictions from LCSR'18~\cite{Zhong:2018exo} and LQCD'19~\cite{Yao:2019vty}.}}
\label{Fig:dg}
\end{figure}

At this stage, we begin to compute the branching fractions of the semileptonic decays $B\to D\ell^+\nu_\ell$ in terms of the $B\to D$ TFF obtained from both the LCSRs and MC approaches. The CKM matrix elements $|V_{cb}|$ is a crucial input parameter in computation, which is taken as $|V_{cb}|=(41.09\pm1.16)\times10^{-3}$~\cite{BaBar:2023kug}. After considering the $B$-meson lifetime $\tau_B$, the numerical results of ${\cal B}(B\to D\ell^+\nu_\ell)$ are presented in Table~\ref{Tab4:BranchingFraction}. We also plot the differential decay width of the semileptonic decays $B\to D\ell^+\nu_\ell$ in Fig.~\ref{Fig:dg}, where other theoretical and experimental results are shown for comparison, such as the Belle'16 Collaboration~\cite{Belle:2015pkj}, LCSR'18~\cite{Zhong:2018exo}, and LQCD'19~\cite{Yao:2019vty}. Furthermore, we also compare our results with those from theoretical predictions and experimental observations to assess their consistency in Table~\ref{Tab4:BranchingFraction}. With which include results from theoretical groups of pQCD'13~\cite{Fan:2013qz}, LCSR'14~\cite{Fu:2013wqa}, LCSR'17~\cite{Zhang:2017rwz}, as well as from experimental groups CLEO'19~\cite{CLEO:1997yyh}, \textsl{BABAR}'10~\cite{BaBar:2009zxk}, Belle'16~\cite{Belle:2015pkj}, Belle'22~\cite{Belle-II:2022ffa}, \textsl{BABAR}'23~\cite{BaBar:2023kug} and HFLAV~\cite{HeavyFlavorAveragingGroupHFLAV:2024ctg}. Our comparison reveals certain deviations from the results of other groups, which may primarily stem from uncertainties in the input parameters. However, within the estimated error range, these results remain reasonable.

\begin{table}
\footnotesize
\caption{The branching  fraction results for $B^0\to D^-\ell^+\nu_\ell$ and $B^+ \to \bar D^0 \ell^+ \nu_\ell$ are presented here. At the same time, other experimental and theoretical results are used for comparison.}. \label {Tab4:BranchingFraction}
\begin{tabular}{l l l}
\hline
~~~~~~~~~~~~~~~~~~~~~~~&Decay Channel ~~~~~~~&Branching fractions\\
\hline

&$ B^0\to D^-\ell^+\nu_\ell$&$(2.10_{-0.38}^{+0.44})\%$\\
\raisebox {2.0ex}[0pt]{LCSR (This Work)}&$ B^+ \to \bar D^0 \ell^+ \nu_\ell$&$(2.26_{-0.41}^{+0.48})\%$\\[1ex]

&$ B^0\to D^-\ell^+\nu_\ell$&$(2.11\pm0.10)\%$\\
\raisebox {2.0ex}[0pt]{MC (This Work) }&$ B^+ \to \bar D^0 \ell^+ \nu_\ell$&$(2.27\pm0.10)\%$\\[1ex]

&$\bar B^0\to D^+\ell ^-\bar{\nu}$&$(1.87\pm0.15\pm0.32)\%$\\
\raisebox {2.0ex}[0pt]{CLEO'19~\cite{CLEO:1997yyh}}&$B^-\to D^0\ell ^-\bar{\nu}$&$(1.94\pm0.15\pm0.34)\%$\\[1.5ex]

&$\bar B^0\to D^+\ell ^-\bar{\nu}$&$(2.23\pm0.11\pm0.11)\%$\\
\raisebox {2.0ex}[0pt]{\textsl{BABAR}'10~\cite{BaBar:2009zxk}}&$B^-\to D^0\ell ^-\bar{\nu}$&$(2.31\pm0.08\pm0.09)\%$\\[1.5ex]

&$B^0\to D^-\ell^+\nu_\ell$&$(2.15\pm0.11\pm0.14)\%$ \\
\raisebox{2.0ex}[0pt]{\textsl{BABAR}'23~\cite{BaBar:2023kug}}&$B^+ \to \bar D^0 \ell^+ \nu_\ell$&$(2.16\pm0.08\pm0.13)\%$ \\[1.5ex]

Belle'16~\cite{Belle:2015pkj} &$\bar{B}^0\to D^+\ell ^-\bar{\nu}$&$(2.31\pm0.03\pm0.11)\%$\\[1.5ex]

&$ B^0\to D^-e^+\nu _e$& $(1.99\pm 0.04\pm 0.08)\%$\\
\raisebox {2.0ex}[0pt]{Belle'22~\cite{Belle-II:2022ffa} }&$ B^+ \to \bar D^0 e^+ \nu_e$&$(2.21\pm 0.03\pm 0.08)\%$\\[1.5ex]

pQCD'13~\cite{Fan:2013qz}&$\bar{B}^0\to D^+\ell ^-\bar{\nu}_{\ell}$&$(2.03^{+0.92}_{-0.70})\%$\\[1.5ex]

&$ B^0\to D^-\ell^+\nu_\ell$&$(2.18\pm0.12)\%$\\
\raisebox {2.0ex}[0pt]{LCSR'14~\cite{Fu:2013wqa}}&$ B^+ \to \bar D^0 \ell^+ \nu_\ell$&$(2.26\pm0.11)\%$\\[1.5ex]

LCSR'17~\cite{Zhang:2017rwz}&$\bar B^0\to D^+\ell\nu_\ell$&$(2.132\pm0.273)\%$\\[1.5ex]

&$ B^0\to D^-\ell^+\nu_\ell$&$(2.12\pm0.02\pm0.06)\%$\\
\raisebox {2.0ex}[0pt]{HFLAV~\cite{HeavyFlavorAveragingGroupHFLAV:2024ctg}}&$ B^+ \to \bar D^0 \ell^+ \nu_\ell$&$(2.21\pm0.02\pm0.06)\%$\\

\hline
\end{tabular}
\end{table}

In addition, in order to provide a new result about $|V_{cb}|$ in the QCD LCSRs and the MC methods, we extract the CKM matrix element $|V_{cb}|$ based on the branching fractions $ B^+ \to \bar D^0 \ell^+ \nu_\ell$ and $ B^0\to D^-\ell^+\nu_\ell$ in \textsl{BABAR}'23~\cite{BaBar:2023kug} and Belle'22~\cite{Belle-II:2022ffa}, respectively. The result is as follows:

\begin{eqnarray}
&& |V_{cb}|_{\rm SR}~(B^+ \text{-Channel})=40.54_{-3.80}^{+4.33}\times 10^{-3},
\\
&& |V_{cb}|_{\rm MC}~(B^+ \text{-Channel})=40.46_{-1.38}^{+1.40}\times 10^{-3},
\\
&& |V_{cb}|_{\rm SR}~(B^0 \text{-Channel})=41.55_{-4.50}^{+4.91}\times 10^{-3},
\\
&& |V_{cb}|_{\rm MC}~(B^0 \text{-Channel})=41.47_{-2.66}^{+2.55 }\times 10^{-3}.
\end{eqnarray}

For comparison, we have collected the results of the CKM matrix element $|V_{cb}|$ from both theoretical and experimental studies. The results are presented in Table~\ref{Tab:IV}, where it can be seen that our predictions for the CKM matrix element $|V_{cb}|$ agree well with other theoretical and experimental results.

\begin{table}
\footnotesize
\caption{Our prediction of $|V_{cb}|$ obtained from the decay channels $ B^+ \to \bar D^0 \ell^+ \nu_\ell$ and $ B^0\to D^-\ell^+\nu_\ell$, compared with other theoretical and experimental  results of various channels (in unit $10^{-3}$).}.\label {Tab:IV}
\begin{tabular}{lllll}
\hline
$|V_{cb}|$~~~~~~~~~~~~~~~~~~~~&$B^+$-Channel~~&$B^0$-Channel\\
\hline
LCSR (This Work) &$40.54_{-3.80}^{+4.33}$ &$41.55_{-4.50}^{+4.91}$\\
MC (This Work) &$40.46_{-1.38}^{+1.40}$ & $41.47_{-2.66}^{+2.55 }$ \\
\textsl{BABAR}'10~\cite{BaBar:2009zxk} &$38.47\pm0.90$ &$38.22\pm0.90$ \\
\textsl{BABAR}'23~\cite{BaBar:2023kug} &$38.67\pm1.41$ & $40.02\pm1.76$ \\
Belle'16~\cite{Belle:2015pkj} &$40.83\pm1.13$ & $40.83\pm1.13$ & \\
Belle'22 ~\cite{Belle-II:2022ffa}& $38.28\pm 1.16$ &$38.28\pm 1.16$ \\
LCSR'14~\cite{Fu:2013wqa}&$41.28_{-4.82}^{+5.68}$ & $40.44_{-4.72}^{+5.56}$\\
LCSR'22~\cite{Gao:2021sav}&$40.2_{-0.5}^{+0.6}|_{\rm th}\pm1.4|_{\rm exp}$ & $40.9_{-0.5}^{+0.6}|_{\rm th}\pm1.0|_{\rm exp}$\\
LQCD'19~\cite{Yao:2019vty}&$41.01\pm0.75$ & / \\
FNAL/MILC'15~\cite{MILC:2015uhg}&$39.6\pm1.7\pm0.2$ & / \\
HPQCD'16~\cite{Na:2015kha}&/ & $40.2\pm0.017\pm0.013$ \\
LQCD'22~\cite{Martinelli:2021onb}&$41.0\pm1.2$ & $41.0\pm1.2$ \\
HQET'16~\cite{Bigi:2016mdz}&$40.99\pm0.97$ &$40.99\pm0.97$ \\
PDG~\cite{ParticleDataGroup:2024cfk} &$41.1\pm1.2$ & $41.1\pm1.2$ \\
HFLAV~\cite{HeavyFlavorAveragingGroupHFLAV:2024ctg} &$38.9\pm0.7$ & $38.9\pm0.7$ \\
\hline
\end{tabular}
\end{table}

\section{Summary}\label{Sec:4}
In this paper, we utilize the MC method in combination with the LCSRs method to investigate the key component of $B\to D$ semileptonic decay, specifically the $B\to D$ TFFs. Firstly, the experimental data points for TFF ${\cal G}(w)$ both from \textsl{BABAR} and Belle Collaborations were collected. It is shown in Fig.~\ref{Fig:G}. We give the TFF ${\cal G}(w)$ behaviors obtained from simulating these data points using the MC method for five separate experiments. It is shown in Fig.~\ref{Fig:fq+}. After considering all the data points comprehensively, we employed the MC method to fit TFF $f_+^{BD}(q^2)$ according to Eq.~\eqref{eq:fq+}, therefore obtaining the relevant data. It is shown in Table~\ref{table:II}. We then took the $f_+^{BD}(q^2)$ calculated by the LCSR method as the fitting object for our research. Finally, we successfully determined the result of the parameter $B_{2;D}$ for the $D$-meson leading-twist LCDA and presented it in Table~\ref{B_results}. Subsequently, within the light-cone range, we provided the graphs of TFF $f_+^{BD}(q^2)$ obtained by the MC and LCSR methods, which are shown in Fig.~\ref{Fig:fp}. Meanwhile, the specific behavior of the $D$-meson LCDA $\phi_{2;D}(x,2\rm GeV)$ is given in Fig.~\ref{Fig:DA1}. The prediction results of the KLS model, LLZ model, LM model, LFQM model, and LCHO model are also included.

Then, we used the SSE method to extrapolate $f_+^{BD}(q^2)$ to the entire $q^2$-region and compared it with the other theoretical and experimental prediction results shown in Fig.~\ref{Fig:fp1}. Meanwhile, the TFFs of $B\to D$ at the large recoil point are presented in Table~\ref{Tab:II}. To evaluate the rationality of our extrapolated results compared with those obtained by the  LCSR, we provided the coefficients of the expansion of $\beta_{k}$ when $k = 2$ in Eq.~\eqref{FSSE} and listed them in Table~\ref{tab:FSSE}. Based on the results in Table~\ref{tab:FSSE}, we presented the specific data of TFF $f_+^{BD}(q^2)$ at different $q^2$-values in Table~\ref{tab:Fq2}. Through analysis, it is found that the error $\Delta<0.00293\%$, which clearly indicates that the results of the SSE extrapolation are in high agreement with those calculated by the LCSR, fully demonstrating the extremely high feasibility of this method. In addition, this paper also presents the variation of another form of TFF, $\mathcal{G}(\omega)$, which is shown in Fig.~\ref{Fig:Gw}. We compared it with other experimental data and theoretical predictions, and the results are in good agreement with them.

Finally, we calculated the differential and total decay width, branching fractions for semileptonic decay $B^+ \to \bar D^0 \ell^+ \nu_\ell$, $ B^0\to D^-\ell^+\nu_\ell$. The values are given in Table.~\ref{Tab4:BranchingFraction}, which also include the numerical results from theoretical groups such as LCSRs, PQCD, as well as experimental groups like \textsl{BABAR}, Belle and CLEO Collaborations. Our predicted branching fractions are consistent with the experimental data within errors. And then, we predict the CKM matrix element $|V_{cb}|$ and presented it in Table~\ref{Tab:IV}. Upon comparison and analysis, it was found to be highly consistent with both theoretical and experimental predictions.
\\

\section{Acknowledgments}
We are grateful to referee for the valuable and useful comments. Hai-Bing Fu and Tao Zhong would like to thank the Institute of High Energy Physics of Chinese Academy of Sciences for their warm and kind hospitality. This work was supported in part by the National Natural Science Foundation of China under Grants No.12265010, 12265009, the Project of Guizhou Provincial Department of Science and Technology under Grants No.MS[2025]219, No.CXTD[2025]030, No.ZK[2023]024.

\end{document}